# Appcessory Economics:
# Enabling loosely coupled hardware / software innovation

Koen Holtman
Philips Research, Eindhoven, The Netherlands
Koen.Holtman@philips.com

**Abstract**

*An appcessory (app + accessory) is a smart phone accessory that is combined with a specially written app to perform a useful function. An example is a toy helicopter controlled by a smart phone app: the full value proposition involves both new hardware outside the phone and new software running inside the phone. Like the smart phone itself and like a PC, the appcessory hardware is a platform: it has the property that it becomes even more valuable if innovative new software is written for it. It is technically possible that innovation in the appcessory space becomes loosely coupled, with different parties specializing to innovate on either the hardware (accessory) side or the software (app) side. However, we are not seeing this loose coupling yet in the current market. We argue that the reason for this is economical: a market framework in which loosely coupled software parties are adequately rewarded is not yet in place. We describe a technical solution for creating the needed market framework, consisting of an automated micropayment system by which money can flow from appcessory hardware makers to app software innovators. This solution also has a broader applicability to the problem of innovation in the Internet of Things.*

**Keywords**: appcessories, smart phones, apps, accessories, micropayments, security, Internet of Things, economics of innovation.

## 1. Introduction

An appcessory (app + accessory) can be defined as a smart phone accessory that is combined with a specially written app to perform a useful function. An example is a toy helicopter controlled by a smart phone app: the full value proposition involves both new hardware outside the phone and new software running inside the phone. A normal smart phone accessory, e.g. a Bluetooth headset or a custom cover does not need any new software in order to be valuable. Like a smart phone and a PC, the appcessory hardware is a platform: it has the interesting property that it becomes even more valuable if innovative new software is written for it. It is technically possible that innovation in the appcessory space becomes loosely coupled, with different parties specializing to innovate on either the hardware (accessory) side or the software (app) side. However, we are not seeing this loose coupling yet in the current market. We argue that the reason for this is economical. A market framework in which loosely coupled software parties are adequately rewarded for their innovation is not yet in place. We describe a technical solution for creating the needed market framework, consisting of an automated micropayment system by which money can flow from accessory hardware makers to app software innovators.

Our technical solution is an automated reward system, using (virtual) coupons, each coupon representing a monetary value, stored in an accessory (inside the hardware part of an appcessory). The hardware part of the accessory will send these coupons to apps that use it, with the apps sending these coupons to a cashing server, so that the bank account of the app author is credited with the coupon value. If there are coupons inside a newly launched accessory, this creates strong incentives for app authors to write apps for this accessory. Spending money on coupons will be much more effective for an accessory vendor to promote his accessory than e.g. spending money on advertising. A further advantage of the proposed system is that it can make timely and fine-grained information available, to both accessory vendors and app makers, about how their products are combined in the market to deliver appcessory value propositions. This greater transparency can be used by both sides to fine-tune further innovation strategies. They thus have a greater chance of starting up a virtuous cycle in which their innovations attract more and more users to a particular appcessory proposition, leading to greater hardware sales, leading to more money available overall to fund further innovation.

## 2. Problems faced by an accessory vendor

We now consider the problem faced by an accessory vendor who believes that he has designed a great new accessory which (in this example) can enhance game play on smart phones. This vendor lacks the necessary in-house skills, or the money, to create many good game apps for the accessory. In parallel with the first sales of this accessory to end users, the accessory vendor needs to get the attention of some game app developers, and convince them to support the accessory in their games, even though the installed base of the accessory is still small. A well-known solution to this problem would be to make up-front payments to some game app developers to add support for the accessory. However, this is problematic in the smart phone world, where there are hundreds of thousands of app authors, and where games that are completely unknown to 99% of the world can be huge for a 1% niche audience. The accessory vendor runs a high risk that he fails to find and fund that particular app author who would be capable of having the golden out-of-the-box idea that would make the combination of this app and this accessory into something that everybody will want to buy. A better solution to this search process is needed. We propose to solve this search problem, to some extent, by transferring it from the accessory vendor to the app makers.

We envisage an automated system that rewards app makers post-facto for proven accessory sales they generated, rather than paying them up front for uncertain results. App authors can search the accessory market (e.g. via gadget web sites) for accessories that might create a creative spark in their minds, and they will get rewarded if this creative spark turns out to be golden.

# 3. Proposed system

In general terms, the proposed system works using (virtual) coupons, where each coupon is a (cryptographically protected) unique object that represents a certain monetary value. When producing a batch of new accessories, say the first batch of 5000 units for marketing via web stores connected to gadget web sites, the accessory vendor embeds 500 'coupons', each worth $0.01, inside each accessory.

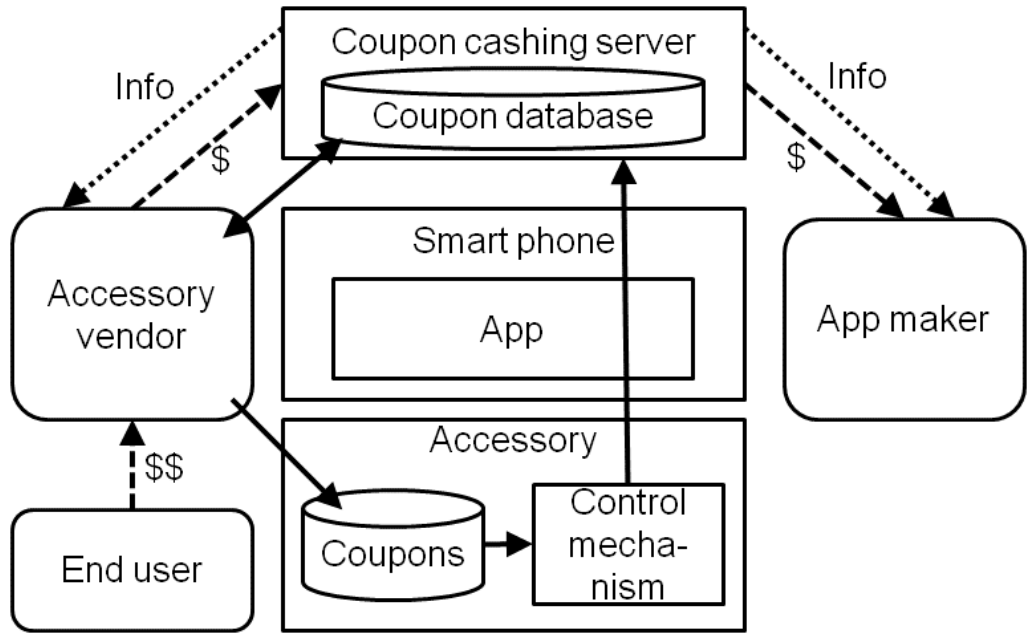

*Figure 1: General system principles. The solid arrows show coupon flow: the accessory vendor creates coupons and stores them in an accessory. The coupons are also registered with a coupon caching server, so that the server can recognise valid coupons. The control mechanism in the accessory monitors its use, and sends coupons to the app as a reward for the app using it. The app forwards the coupons to the coupon caching server, tohether with the identity of the app maker, so that the app maker can be rewarded with the coupon value. The dashed lines show the money flow: end users pay to buy accessory hardware, part of this money is kept by the accessory vendor, and another part goes, via the coupon cashing server, to the app makers. The dotted lines show the information gathered by the coupon cashing server flowing to the accessory vendors and the app makers, allowing them to optimise further actions.*

A control mechanism inside the accessory can detect when the accessory is being used by an app. If the accessory is being used, the mechanism occasionally sends a coupon to the app, over the same data connection that the app is using to control the accessory. For example, the control mechanism is programmed to send one coupon for every minute of use, until the coupons run out. The app, which has been programmed to expect getting coupons from the accessory, is programmed to forward each coupon to a 'coupon cashing server' on the internet, together with an ID number that identifies the app maker. The coupon caching server is programmed to verify that the coupon is real, and that is has not yet been used (sent) before. If it is real and sent for the first time, the server will credit (the bank account of the) app maker with the amount of money represented by the coupon, money that is debited to the accessory vendor, who will also likely have to pay a service fee. Figure 1 illustrates the system.

The effect of the above setup is as follows. When a customer buys and uses an accessory, then the makers of the game apps that the customer uses to control the accessory will be rewarded, at least if these apps are used within the first 500 minutes (8.3 hours) of use of the accessory. This means that exactly those game app authors get rewarded that were most likely responsible for the user buying the accessory in the first place.

The accessory vendor can choose to change the number of coupons (and the value of the coupons) put in each subsequently produced batch of accessories, initially, the embedded coupons might represent a lot of value, maybe even a number close to the sales price of the accessory itself. If the accessory becomes moderately successful, the accessory vendor can chose to lower the value of the coupons to be put in further production batches, relying on other mechanisms (e.g. media coverage, good product reviews on gadget web sites) to drive further sales.

This system allows the accessory vendor, if he so wishes, to commit large amounts of money to an attempt to jump-start a virtuous cycle. The up-front design, production, and certification costs of an electronic accessory can easily be in the order of $50.000, even if the initial production run is kept small. Thus, putting $20 worth of coupons in each accessory from an initial 1000-unit production run will not represent an insurmountable extra cost to the accessory vendor. This $20.000 can be an appealing bounty however to app makers – especially app makers in the 'professional hobbyist' class. (It has been reported [1] that the cost of developing an app by a 'day job' professional is about $15.000 - $30.000, depending on factors like platform, cost of living (location) of the developer, type of app, etc.) To prevent abuse of the system, it is unwise for an accessory vendor to embed coupons in an accessory that are worth more than the accessory retail price.

The coupon cashing server can also provide the accessory vendors detailed reports of which apps were used (in the coupon-generating period) with their accessories, and the app authors detailed reports on which accessories have been most rewarding for them to support. They can then use this information to optimize their activities, leading to more efficient innovation overall. The server might also produce reports showing which accessories in the market (either sold but not used much yet, or still lying unsold in the stores) contain the higher number of unredeemed coupons, pointing app makers into the direction of the most rewards. The reports produced by the coupon cashing server should anonimize smart phone end user identities, even though they might be technically knowable to the server, in order to address privacy concerns. Conceivably, an opt-in or opt-out mechanism is in place.

To deal with apps that might use the accessory, but that were not designed to handle the redeeming of coupons, it may be a requirement that the app must first inform the control mechanism inside the accessory that it is willing to receive coupons.  If no such signal is received, the control mechanism will not emit any coupons: this avoids the coupons from going to waste.

## 4. Cryptographic details and implementation cost considerations

We propose that coupons are large binary numbers, of say 128 bits.  This makes it useless for an attacker to send randomly invented coupon numbers to the coupon cashing server in the hope of redeeming an actually existing coupon.  To keep implementation costs of the accessory low, we propose that the accessory contains a 128-bit 'coupon key' (e.g. realized using flash memory), a random number that is different for each accessory, and it is set in the factory.  The accessory also contains a 'coupon counter' (e.g. realized as flash memory, so it keeps its value even if the power goes off), that is set to 0 in the factory, and increased whenever a coupon is generated.  A 'coupon max' number is set in the factory to the number of coupons that should be generated.  The generator encrypts the value of the coupon counter using the coupon key, e.g. using AES128.  The result is the coupon.

In the factory, the coupon key is also used, inside the factory equipment, to generate all coupons that the accessory will ever emit. These coupons are then sent to the coupon cashing server, together with information of how much each coupon is worth.  The server adds these coupons and this data to a relational database table inside itself (and well protected from hackers), with each entry in the table being of the form (coupon number, worth, accessory vendor id, redeemed true/false), where the table is indexed to support lookups on coupon number.  When a coupon number is received from an app, the server will do a database lookup to retrieve the coupon entry.  If it is not found, the server assumes an attack and no money is transferred.  If it is found, and the coupon has not yet been redeemed, money is booked for transfer, and the coupon status is changed to redeemed.

In very low cost accessories that do not have on-board flash memory, the coupon key might be stored using on-chip 'fuses' that are blown in the factory to represent the number.  In that case, the coupon counter need not be stored in the accessory: the accessory is instead set up to store the coupon counter value outside it, e.g. inside the coupon caching manager described below, or on the coupon cashing server.

As an alternative to storing a coupon counter, the coupon generator could be set up to generate a random number, between zero and 'coupon max'-1, whenever a coupon is requested, and to encrypt this number.  This will lead to the duplicate generation of coupons, but these duplicates can be filtered out by the coupon cashing server.  An effect of this approach will be that the chance of an app getting valuable coupons out of the accessory will slowly get lower over time, rather than dropping to zero suddenly.

## 5. The coupon cashing manager

The integrity of the system might be attacked as follows. Modern smart phones can run multiple apps at the same time, and this allows the malicious author of a e.g. a calendaring app, that stays on all the time, to include code in his app that searches for connected accessories, and starts using them, unbeknownst to the end user of the phone, in the hope of collecting coupons. Such use will be of no benefit to the user of course, nor will it help the accessory vendor to create more sales. To prevent this abuse, there are multiple options. First, if actuators are present in the accessory, e.g. the accessory is a lamp, the control mechanism is configured to monitor the use of the actuators and generate coupons only when these actuators are controlled, e.g. when the light setting is changed. To get coupons out of the accessory, the calendaring app will then have to make the accessory behave in a way that will look suspicious to the end user.

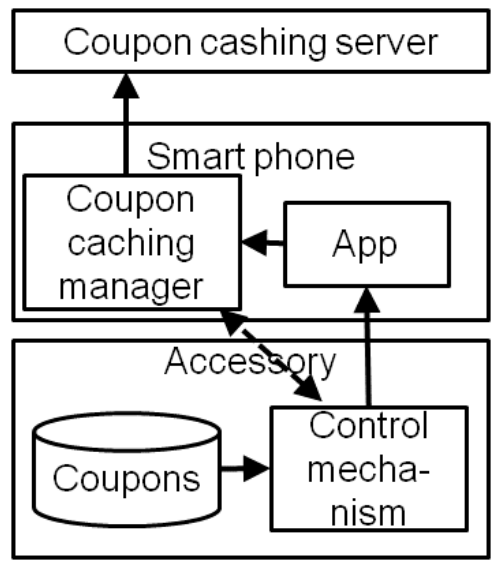

*Figure 2: Use of a coupon caching manager, a special app in the smart phone, as an additional security measure to protect the integrity of the rewards system against e.g. malicious app authors. By insterting itself in the coupon stream, the manager can detect certain attacks after the fact, and prevent certain other attacks from happening at all.*

As a second option, a special app called a coupon cashing manager is run inside the smart phone, concurrently with the app that uses the accessory. The coupon cashing manager is an app written by the same organization that runs the coupon cashing server, and is able to prove to the server (using e.g. cryptographic credentials) that is it a legitimate coupon cashing manager app that has not been tampered with. The server is set up to refuse accepting coupons sent by apps other than legitimate coupon cashing managers. The app in the smart phone that is actually using the accessory is set up to deliver the received coupons to the coupon cashing manager, which will forward them to the server. The manager also compiles a report, that the user can see, about which app authors got paid how much by which accessories the user has. This report allows an end user to discover abuse: the end user can discover that a calendaring app is collecting coupons maliciously.  The end user can then report this abusing app to the organization running the coupon cashing server, e.g. by using a 'report abuse' button provided by the coupon cashing manager along with its report.  If abuse is reported, the organization running the coupon cashing server can investigate and take measures (like suspending payments and maybe instructing their bank to collect back the money already paid).  The organization can then redistribute the money recovered to legitimate app authors, e.g. based on statistics collected from the smart phones of users where no malicious apps where running.

Depending on the smart phone security system and user preferences, game apps will not get internet access, so they are not even capable of contacting a coupon cashing server directly as shown in figure 2. This is another reason to have the coupon cashing manager.   In an android phone, if an USB accessory is first plugged in, the Android phone can locate an appropriate app recommended by the accessory vendor.  This mechanism can be used by the vendor to recommend that the coupon cashing manager is installed.  The coupon cashing manager can then also have the additional feature of being able give the user a list of popular apps that use the accessory, with popularity scores derived e.g. from statistics obtained by the coupon cashing server.

## 6. Business model alternatives

Depending on the nature of the accessory and the business model of the accessory vendor, the accessory vendor can fine-tune the control mechanism to release coupons in a particular way (e.g. the first 100 coupons in the first hour of use, the next 100 divided over the next 10 hours), or to release more coupons only if certain features of the accessory are used, e.g. distinguishing features that the accessory vendor wants to emphasize.   The coupon generating mechanism, or any of the cashing mechanisms, might also interact with the end user, to allow the end user to express a bias about which apps should be rewarded most.   It is also possible, e.g. by adding extra mechanisms to the coupon cashing server, to create different reward structures than a fixed fee for each coupon.  An app author might be rewarded with $.50 for the first coupon that one of their apps sends, $.25 for the second, $.10 for the third, etc, up to a maximum of $1 per app author, with the server paying out different app authors until the balance of $5 runs out.  This way, the accessory vendor might promote a the availability of a greater mix of apps – it is no longer possible that 100% of the coupon money goes to the author of author of a single hugely popular game app.  The accessory might be configured in this case to generate 5000 coupons during the first 1000 hours of use, with the expectation that the balance will probably run out long before the last coupon is generated.   While this variable pricing technique might not be appealing to accessory vendors who operate on an 'impulse buy' market, it can be very useful to accessory vendors who want to make and keep an implied promise of lasting value for their products: this technique allows such vendors to promote the creation of new apps that give its customers new (game) experiences with an accessory they have already bought.

Several money collection and payment models are possible for the organization running the coupon cashing server. To prevent abuse, the organization might require payment from accessory vendors up front. To prevent abuse at the app author side, the organization can delay payment of credited money for some time, so that any abuse is likely to be reported before payments are made -- this is typical also in the model of how app stores deal with fee payments to app authors for apps that are not free. If payments to app authors are delayed, it is important that they have a way to get a 'live' view of their balance due, so that they can react timely to new events in the marketplace.

## 7. Re-factoring for lower accessory costs

An accessory that has to be produced very cheaply may lack the capacity to run a sophisticated control mechanism for releasing coupons inside it. To enable such accessories, the proposed design can be re-factored to place some or all of the control mechanism functions inside the coupon cashing manager. The coupon cashing manager may in theory relay all interactions between the app and the accessory through itself, acting as a kind of device driver for the accessory, as far as the app is concerned. In an extreme case, accessory only contains the coupon key, or another form of secure ID that uniquely and reliably identifies this particular instance of the accessory to the coupon cashing manager app, to prevent double-counting if a user owns multiple accessories of the same type.

To protect against certain attacks, a security mechanism has to be used in the accessory, so that only a valid coupon cashing manager app is able to obtain the information needed to generate coupons and send them onwards to the coupon cashing server. The security mechanism might work by requiring an interaction with the coupon cashing server – in that case the cryptographic elements needed in the accessory could be minimal. As an extreme minimalist example, the security might consist only of making the unique ID a very large (say 128 bit) random number, that could be read by any app connected to the accessory, but that can only be used once, when sent to the coupon cashing server, to receive the coupon key for the accessory. The coupon cashing server is set up to perform this key granting operation only when contacted by an app for which it can verify that it is a legitimate coupon cashing manager app. Mechanisms that allow servers to verify app legitimacy, with reasonable security, are part of many smart phone OSes.

## 8. Application to the Internet of Things

While we concentrated on the appcessory case in this paper, it should be noted that the proposed system can be applied in other contexts too. Consider an Internet of Things future, in which software production and hardware creation are activities performed by independent specialized entities all trying to make money. In such a future, the proposed system is applicable to promote investments and boost innovation efficiency, leading to a faster creation of product combinations that people want to pay money for to experience.

## 9. Conclusion

If we look at software creation as an economic activity, what is important about the smart phone app revolution is not so much that smart phones are another successful software application platform (like PCs and the Web before it), but that smart phone app stores are a real-world existence proof of a working micropayment system that rewards software creation and innovation.  This micropayment model for supporting software creation stands in contrast to the 'macro-payment' and volunteer 'free/open source' models that are dominant in PC software, and to the advertising-based monetization model that supports most web applications.  When it comes to privacy considerations, a micropayments-funded Internet of Things is preferable, compared to an advertising-funded one.

In this paper we have proposed an innovation model for the appcessory value space that depends on micropayment transactions to app authors, funded out of money reserved by accessory vendors.  The model does not burden the end user with any major mental workload: interested end users may interact with their coupon caching manager app, and they have a role to play in detecting certain types of system abuse, but the uninterested end user will only be faced with having to make economic decisions when shopping for accessory hardware.    The revenue stream from the coupon cashing server will allow app authors to offer their apps as free apps to the end user.

This paper also makes a contribution by considering system cost and security.  Various threats are discussed and we show how these can be addressed without driving up the bill of material costs of the accessory.

It is too early to say whether the appcessory market, or the Internet of Things market for that matter, will really take off.  This paper makes a contribution by proposing an enabling mechanism and working out the technical details.  While the mechanism is interesting in itself, we have as yet found no real means of determining its potential impact as a market enabler.  This is a subject for possible future work.